%% file: main.tex
\documentclass[12pt]{iopart}
\usepackage{iopams}
\usepackage{graphicx}

\input{macros}
\begin{document}
\input{title}
\input{introduction}
\input{qpipeline}
\input{motivations}
\input{clustering}
\input{testing}
\input{conclusion}

\input{acknowledgements}

\section*{References}
\bibliographystyle{unsrt}
\bibliography{bibliography}
\end{document}

%% file: title.tex
\title[Enhancing GW burst search by clustering]
      {Enhancing the capabilities of LIGO time-frequency
       plane searches through clustering}

\author[R Khan, S Chatterji]{R Khan$^1$, S Chatterji$^2$}

\address{$^1$ Columbia Astrophysics Laboratory,
              Columbia University,
              Pupin Labs Rm 1027,
              MC 5247,
              New York, NY 10027 USA}
\address{$^2$ LIGO Laboratory,
              California Institute of Technology,
              MS 18-34,
              Pasadena, CA 91125 USA}

\eads{\mailto{rubab@astro.columbia.edu}, \mailto{shourov@ligo.caltech.edu}}

\input{abstract}

\pacs{04.80.Nn, 07.05.Kf, 95.55.Ym, 95.75.Pq}

\submitto{\CQG}

\maketitle

%% file: abstract.tex
\begin{abstract}

One class of gravitational wave signals LIGO is searching for consists
of short duration bursts of unknown waveforms.  Potential sources
include core collapse supernovae, gamma ray burst progenitors, and
mergers of binary black holes or neutron stars.  We present a
density-based clustering algorithm to improve the performance of
time-frequency searches for such gravitational-wave bursts when they
are extended in time and/or frequency, and not sufficiently well known
to permit matched filtering.  We have implemented this algorithm as an
extension to the QPipeline, a gravitational-wave data analysis
pipeline for the detection of bursts, which currently determines the
statistical significance of events based solely on the peak
significance observed in minimum uncertainty regions of the
time-frequency plane. Density based clustering improves the
performance of such a search by considering the aggregate significance
of arbitrarily shaped regions in the time-frequency plane and
rejecting the isolated minimum uncertainty features expected from the
background detector noise.  In this paper, we present test results for
simulated signals and demonstrate that density based clustering
improves the performance of the QPipeline for signals extended in time
and/or frequency.

\end{abstract}

%% file: introduction.tex
\section{Introduction}
\label{sec:Introduction}

The first generation of interferometric gravitational wave detectors
have now collected data at their design strain
sensitivities~\cite{ref:LIGO_status, ref:LIGO_S5_Sensitivity,
ref:virgo, ref:geo06, ref:ligovirgo, ref:LIGO_NIM_2004}, and an
improved generation of detectors~\cite{ref:EnhancedLIGO,
ref:AdvLIGOSens} is already under development. Even at this
unprecedented level of sensitivity, potentially detectable signals
from astrophysical sources are expected to be at or near the limits of
detectability, requiring carefully designed search algorithms in order
to identify and distinguish them from the background detector noise.
In this study, we focus on the problem of detecting the specific class
of gravitational wave signals known as gravitational-wave bursts
(GWBs).  These are signals lasting from a few milliseconds to a few
seconds, for which we do not have sufficient theoretical understanding
or reliable models to predict a waveform.  This includes signals from
the merger of binary compact objects, asymmetric core collapse
supernovae, the progenitors of gamma ray bursts, and possibly
unexpected sources.

Since accurate waveform predictions do not exist for GWBs, the typical
method to identify them is to project the data under test onto a
convenient basis of abstract waveforms that are chosen to cover a
targeted region of the time-frequency plane, and then identify regions
of this search space with statistically significant excess signal
energy~\cite{ref:excess_power}.  In this study, we focus on one such
burst search algorithm, the QPipeline~\cite{ref:qpipeline}, which
first projects the data under test onto an overlapping basis of
Gaussian enveloped sinusoids characterized by their center time,
center frequency, and quality factor.  A trigger is recorded whenever
this projection exceeds a threshold value, with the magnitude of the
projection indicating the significance of the trigger.  Since the
triggers are considered separately, the existing algorithm currently
under-reports the total energy and true significance of those signals
that are extended in time and/or frequency, since they have a
significant projection onto multiple independent basis functions.
Since GWB signals with such extended features are commonly observed in
simulations of core collapse supernovae, the mergers of binary compact
objects, and instabilities of spinning neutron stars, there are good
reasons to try to improve the sensitivity of the search algorithm to
such sources.

To improve the sensitivity of the QPipeline to signals that are
extended in time and/or frequency, we have investigated extensions to
the QPipeline that also consider the combined statistical significance
of arbitrarily shaped clusters of projections in the time-frequency
plane.  Although a number of clustering algorithms are commonly
available~\cite{ref:clustering_review}, this work focuses on a density
based clustering algorithm due to its ability to also decrease the
false detection probability of GWB searches by rejecting isolated
single projection events associated with noise fluctuations.  In this
paper we present the details of a density based clustering algorithm
implementation as an extension to the QPipeline, and demonstrate the
resulting improved performance of the QPipeline for signals that are
extended in time and/or frequency.

The paper is structured as follows.  Section~\ref{sec:qpipeline}
briefly describes the QPipeline burst search algorithm.
Section~\ref{sec:motivations} considers the motivations for clustering
and surveys some of the available approaches.
Section~\ref{sec:clustering} presents the details of the proposed
density based clustering algorithm.  Section~\ref{sec:testing}
demonstrates the benefit of the proposed approach for detecting
simulated gravitational wave bursts of various waveforms.  Finally, in
section~\ref{sec:conclusions}, we present our conclusions and discuss
possible future investigations.

%% file: qpipeline.tex
\section{The QPipeline burst search algorithm}
\label{sec:qpipeline}

The QPipeline is an analysis pipeline for the detection of GWBs in
data from interferometric gravitational wave
detectors~\cite{ref:qpipeline}.  It is based on the
Q-transform~\cite{ref:qtransform}, a multi-resolution time-frequency
transform that projects the data under test onto the space of Gaussian
windowed complex exponentials characterized by a center time $\tau$,
center frequency $\phi$, and quality factor $Q$.

\begin{equation}
\label{eqn:qtransform}
X(\tau, \phi, Q) = \int_{-\infty}^{+\infty} x(t) \, e^{- 4 \pi^2 \phi^2 (t - \tau)^2 / Q^2}
\, e^{-i 2 \pi \phi t} \, dt
\end{equation}

The space of Gaussian enveloped complex exponentials is an overlapping
basis of waveforms, whose duration $\sigma_t$ and bandwidth
$\sigma_f$, defined as the standard deviation of the squared Gaussian
envelope in time and frequency, have the minimum possible
time-frequency uncertainty, $ \sigma_t \sigma_f = 1 / 4 \pi$, where $Q
= \phi / \sigma_f$.

There is good reason to select an overlapping basis of
multi-resolution minimum-uncertainty functions.  Absent detailed
knowledge of the gravitational waveform, such a basis provides the
tightest possible constraints on the time-frequency area of unmodeled
signals, permitting the time-frequency distribution of signal energy
to be non-coherently reconstructed while incorporating as little noise
energy as possible.  A choice of basis that does not have minimum
time-frequency uncertainty would typically include more noise than
necessary, decreasing signal to noise ratio.  The exception would be a
restricted search for a known set of waveforms, in which a matched
filter search, where the template matches the target signal, would be
optimal.  This type of restricted search is not the focus of this
paper.  Another benefit of the tighter time-frequency constraints
afforded by a multi-resolution sine-Gaussian template bank is the
decreased likelihood for false coincidences, when testing for
coincidence between multiple detectors.

In practice, the Q transform is evaluated only for a finite number of
basis functions, also referred to here as templates or tiles.  These
templates are selected to cover a targeted region of signal space, and
are spaced such that the fractional signal energy loss $-\delta Z/Z$
due to the mismatch $\delta \tau$, $\delta \phi$, and $\delta Q$
between an arbitrary basis function and the nearest measurement
template,

\begin{equation}
\label{eqn:mismatch}
\frac{-\delta Z}{Z} \simeq
\frac{4 \pi^2 \phi^2}{Q^2} \, \delta \tau^2 +
\frac{2 + Q^2}{4 \phi^2} \, \delta \phi^2 +
\frac{1}{2 Q^2} \, \delta Q^2 -
\frac{1}{\phi Q} \delta \phi \, \delta Q,
\end{equation}
is no larger than $\sim\!\!20\%$.  This naturally leads to a tiling of
the signal space that is logarithmic in Q, logarithmic in frequency,
and linear in time.

The statistical significance of Q transform projections are given by
their normalized energy $Z$, defined as the ratio of squared
projection magnitude to the mean squared projection magnitude of other
templates with the same central frequency and $Q$.  For the case of
ideal white noise, $Z$ is exponentially distributed, and related to
the matched filter SNR $\rho$~\cite{ref:matched_filter} by the
relation

\begin{equation}
Z = |X|^2 / \langle |X|^2 \rangle_{\tau} = - \ln P(Z^{\prime} > Z) = \rho^2 / 2.
\end{equation}

The QPipeline consists of the following steps.  The data is first
whitened by zero-phase linear predictive
filtering~\cite{ref:makhoul,ref:qpipeline}.  Next, the Q-transform is
applied to the whitened data, and normalized energies are computed for
each measurement template.  Templates with statistically significant
signal content are then identified by applying a threshold on the
normalized energy.  Finally, since a single event may potentially
produce multiple overlapping triggers due to the overlap between
measurement templates, only the most significant of overlapping
templates are reported as triggers.  As a result, the QPipeline is
effectively a templated matched filter
search~\cite{ref:matched_filter} for signals that are Gaussian
enveloped sinusoids in the whitened signal space.

\begin{figure}[!t]
\begin{center}
\scalebox{1.0}[1.45]{\includegraphics[angle=0,width=75mm]{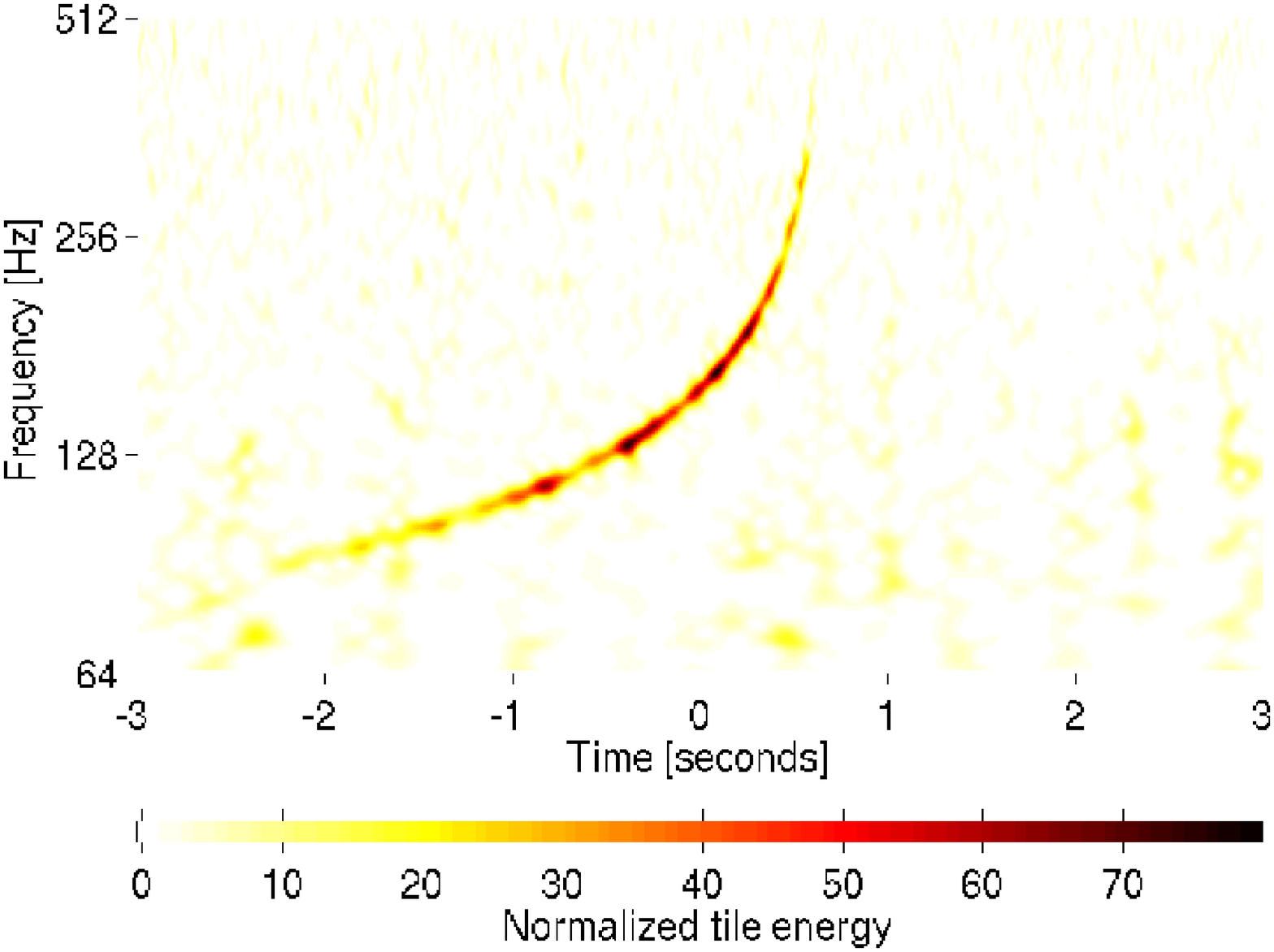}}
\scalebox{1.0}[1.45]{\includegraphics[angle=0,width=75mm]{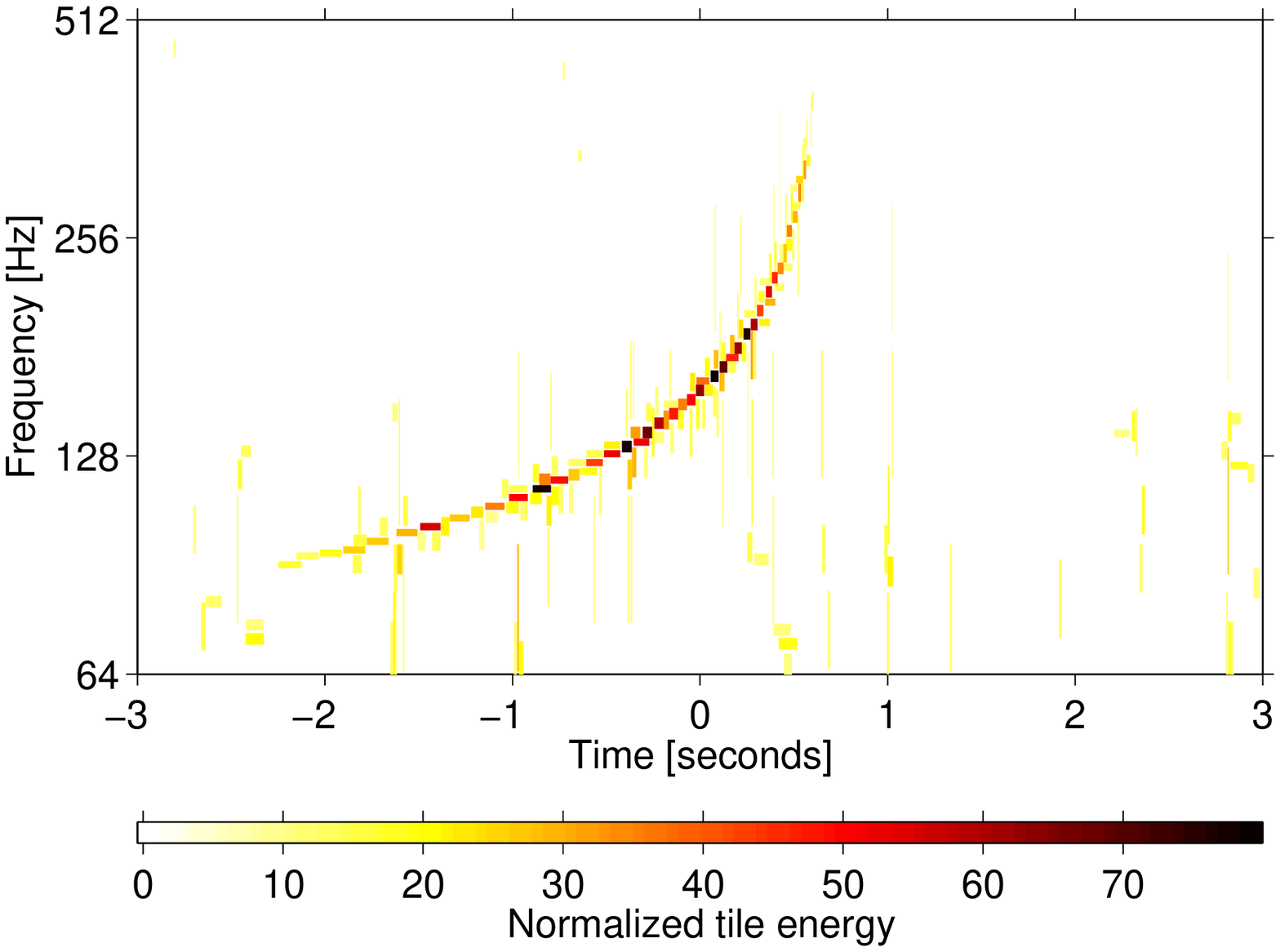}}

\end{center} \caption{The QPipeline view of the inspiral phase of a
simulated optimally oriented 1.4/1.4 solar mass binary neutron star
merger injected into typical single detector LIGO data with an SNR of
48.2 as measured by a matched filter search targeting inspiral signal.
The QPipeline projects the whitened data onto the space of time,
frequency, and $Q$.  The \textit{left panel} image shows the resulting
time-frequency spectrogram of normalized signal energy for the value
of Q that maximizes the measured normalized energy, while the
\textit{right panel} image shows the time-frequency distribution of
only the most significant non-overlapping triggers regardless of $Q$.

The authors gratefully acknowledge the LIGO Scientific Collaboration hardware
injection team for providing the data used in this figure.}
\label{fig:example}
\end{figure}

Figure~\ref{fig:example} shows an example of the QPipeline applied to
the inspiral phase of a simulated binary neutron star coalescence
signal injected into typical single detector LIGO data with an SNR of
48.2.

%% file: motivations.tex
\section{Motivations and options for clustering}
\label{sec:motivations}

Currently, the QPipeline considers the significance of triggers
independently.  The detectability of a particular GWB signal therefore
depends upon its maximum single projection onto the space of Gaussian
enveloped sinusoids.  As a result, the QPipeline is most sensitive to
signals with near minimum time-frequency uncertainty, and less
sensitive to signals that are extended in time and/or frequency such
that their energy is spread across multiple non-overlapping basis
functions.  For example, the detectability of the simulated inspiral
signal shown in Figure~\ref{fig:example} is currently determined by
the single most significant tile near the center of the signal, which
has a single tile SNR of 12.7.  This is significantly less than the
SNR of 48.2 that is recovered by a matched filter search tuned for
this waveform.

Although the above example focuses on binary neutron star inspiral
waveforms, matched filter search methods are more appropriate to
search for such signals, since there waveform is well
understood~\cite{ref:inspiral1,ref:inspiral2,ref:inspiral3,ref:inspiral4}.
We focus on them here only because they represent an astrophysically
interesting example case of a signal which is extended in time and
frequency.  The QPipeline is not intended to search for inspiral
signals, but instead focuses on the search for other transient sources
such as the less well understood merger phase of coalescing binary
compact objects, core collapse supernovae, and instabilities in
spinning neutron stars, many of which are also expected to produce
waveforms that are extended in time and/or
frequency~\cite{ref:mergers,ref:supernovae,ref:Ott,ref:Dimmelmeier}.
As a result, we seek a method to improve the sensitivity of the
QPipeline to signals that are extended in time and/or frequency that
is applicable to the general case of astrophysically unmodeled bursts,
and is not specific to any one particular waveform.

An obvious solution is to simultaneously consider the aggregate
significance of all tiles that comprise the signal.  This requires an
approach that identifies clusters of related tiles in the
time-frequency plane.  In this context, we define clustering as the
grouping of the set of all significant QPipeline tiles into subsets,
such that all tiles within a subset are closely related by their
relative distance in the time-frequency plane.

There are many different clustering methods available; however, they
all tend to fall into three categories~\cite{ref:clustering_review}.

\begin{figure}[!t]
\begin{center}

{\includegraphics[angle=0,width=75mm]{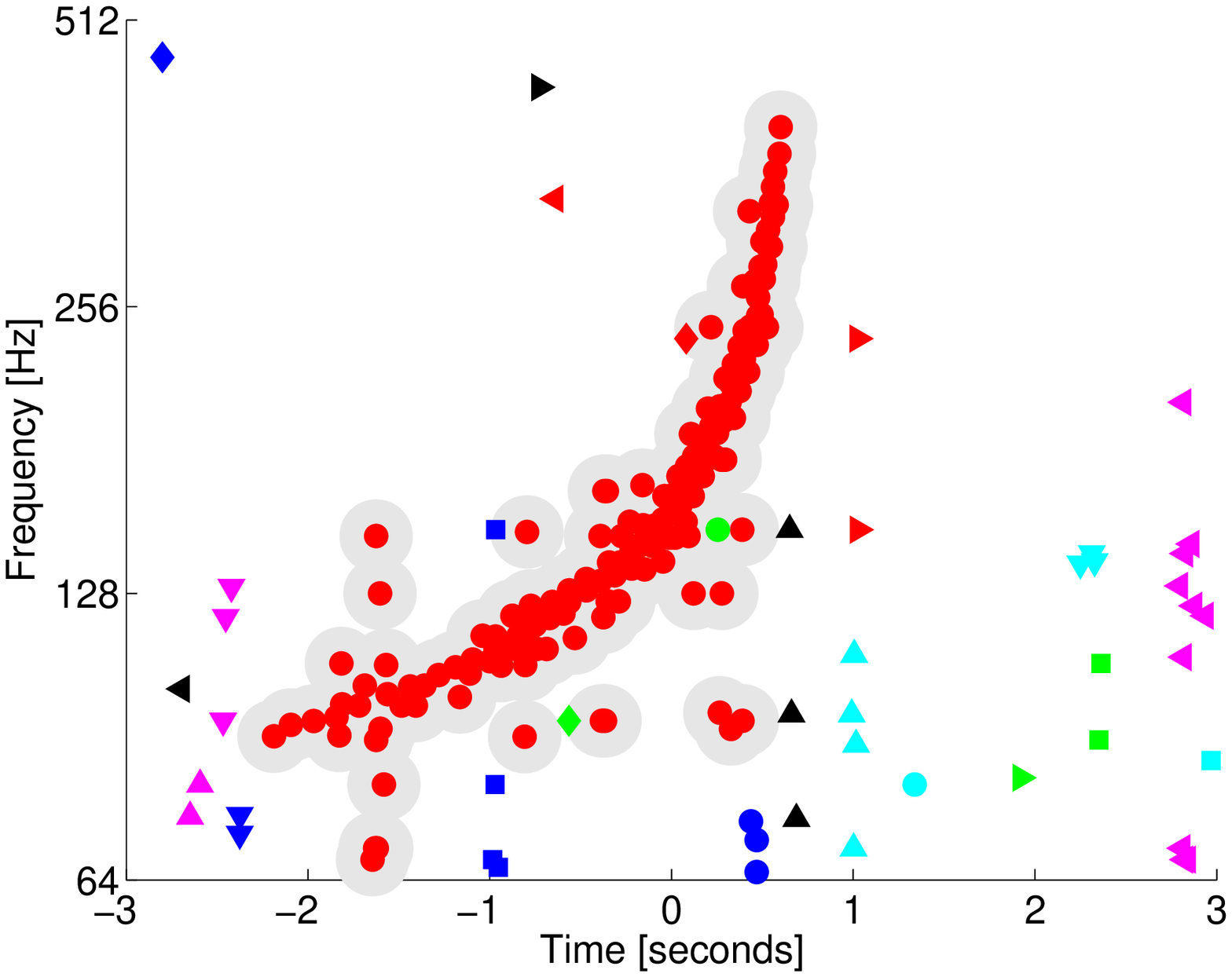}}
{\includegraphics[angle=0,width=75mm]{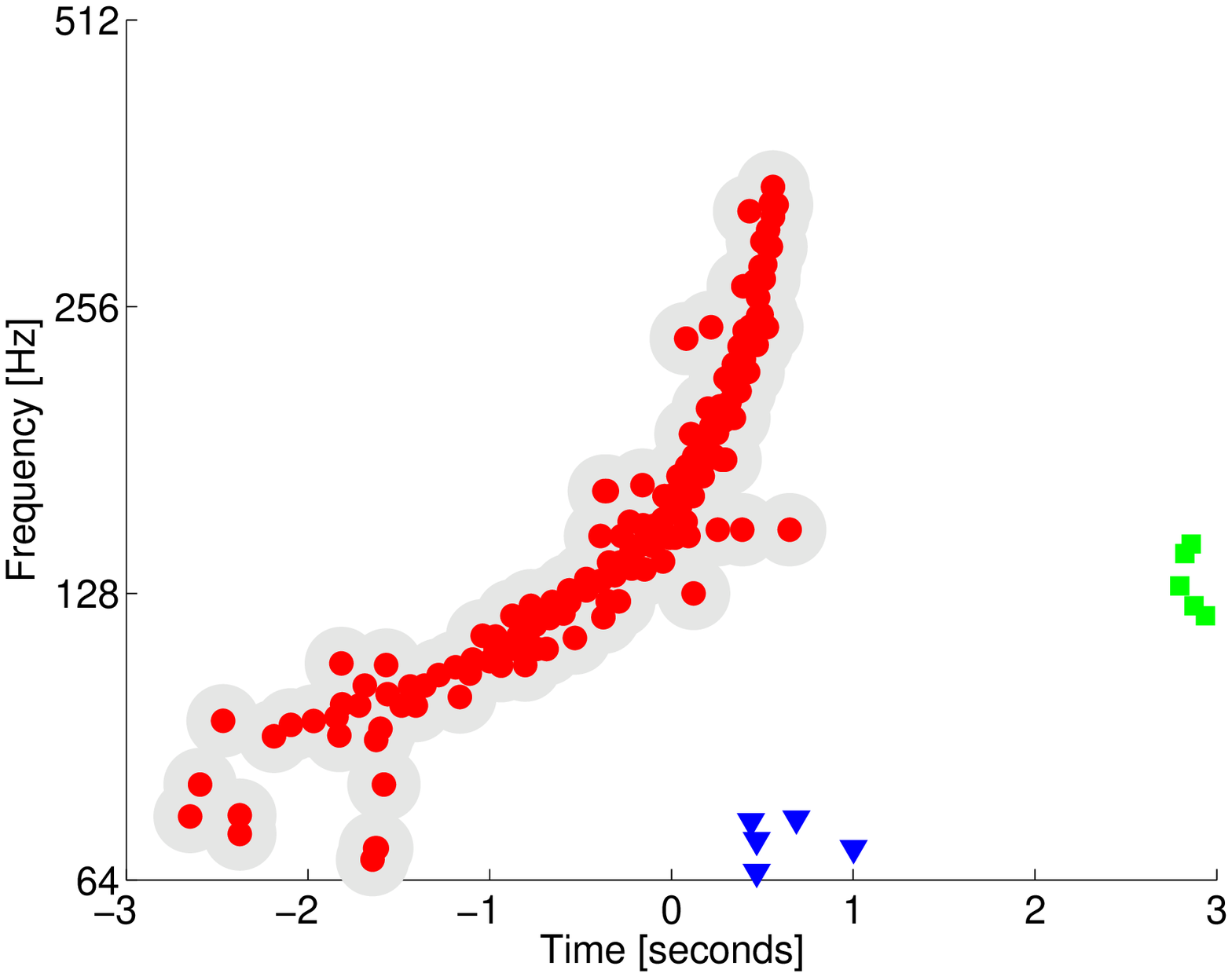}}
\end{center}
\caption{Two different clustering methods were applied to the
QPipeline triggers from Figure~\ref{fig:example}.  The \textit{left
panel} image shows the result of applying a hierarchical based
clustering method~\cite{ref:clustering_review}, while the
\textit{right panel} image shows the result of applying the proposed
density based clustering method~\cite{ref:dbcluster}.  Here each
combination of color and shape denotes a different cluster.
Although both hierarchical and density based clustering approaches
succeed in isolating most of the signal energy within a single
cluster, the density based approach has the additional advantage of
discarding isolated triggers due to the background detector noise.}
\label{fig:example_cont}
\end{figure}

{\em Partitioning methods} The classical example of a partitioning
method is the K-means algorithm~\cite{ref:kmeans}.  In K-means
clustering, a fixed number of clusters presupposed, and an initial
guess to partition objects into these clusters is made.  A centroid is
computed for each cluster, and the total sum distance of all objects
from their cluster centroid is computed as a figure of merit.  K-means
iteratively reassigns objects to different clusters until this figure
of merit is minimized.

There are a number of drawbacks to the K-means approach, as described
in~\cite{ref:kmeans,ref:clustering_review}.  The first is that it
presupposes a fixed number of clusters, though there are variations
that allow the number of clusters to change~\cite{ref:smeans}.  The
other difficulty with the K-means approach is the tendency to produce
spherical or ellipsoidal clusters rather than more complicated
arbitrary shapes.  This is acceptable for some signal morphologies but
not so in general.  For example, the signal expected from inspiralling
binary compact objects, which is long and extended in time and
frequency, would not be easily identified by K-means
clustering~\cite{ref:smeans}.  A third difficulty with the K-means
approach is the sensitivity to the initial guess, and the possibility
of the algorithm to identify a local rather than global
minimum~\cite{ref:clustering_review}.

{\em Hierarchical methods} This type of clustering algorithm first
evaluates the pairwise difference between all $N$ objects, then arranges
them into a tree structure, where each object to be clustered is a
leaf~\cite{ref:Hierarchical_1}.  In the agglomerative hierarchical
approach, the tree is constructed in $N$ levels.  At each level, the
closest pair of leaves and/or branches is merged, with the $N$th level
representing a cluster of all objects. A cluster is formed by cutting
this tree based on some criteria such as a maximum distance or the
inconsistency between the distance between cluster leaves or branches
and the distance to the next closest leaf or branch.

Hierarchical clustering has the flexibility of producing arbitrary
numbers of clusters with arbitrary shape, and is therefore more
applicable to the problem of clustering in time, frequency, and Q for
gravitational-wave burst detection.

The left panel from Figure~\ref{fig:example_cont} shows an example of
hierarchical clustering, implemented using the functions ``linkage''
and ``cluster'' from MATLAB Statistics Toolbox ~\footnote{Statistics Toolbox Software Version
$5.2 (R2006a)$;

$http://www.mathworks.com/access/helpdesk/help/toolbox/stats/rn/bqmfe_z-10.html$
} applied to the detection of a simulated binary neutron star inspiral
signal. The draw back to this approach is that it presupposes that
every data point must be included in one cluster or another, even if a
cluster is to be constituted of only one data point. As a result, it
produces a large number of insignificant clusters and tends to build
clusters of unrelated data points.

{\em Density based methods} Density based
clustering~\cite{ref:dbcluster} is a variation on the hierarchical
clustering approach.  Instead of constructing a tree structure,
density based clustering starts with single object and iteratively
adds objects to that cluster using a predefined set of criteria based
on the density of objects within a given neighborhood radius, until
the criteria is no longer met and all objects have been tested.

Like other hierarchical methods, density based methods also allow an
arbitrary number of clusters as well as arbitrary cluster shape.  They
also have the advantage of rejecting single isolated data points that
are not potentially related with a large number of points. Thus this
method only produces clusters with multiple data points and can
successfully exclude unrelated points from a cluster. For this reason,
we have focused on density based methods in this work.

%% file: clustering.tex
\section{Density based clustering algorithm}
\label{sec:clustering}

\begin{figure}[!t]
\begin{center}
\scalebox{0.95}[0.95]{\includegraphics[angle=0,width=80mm]{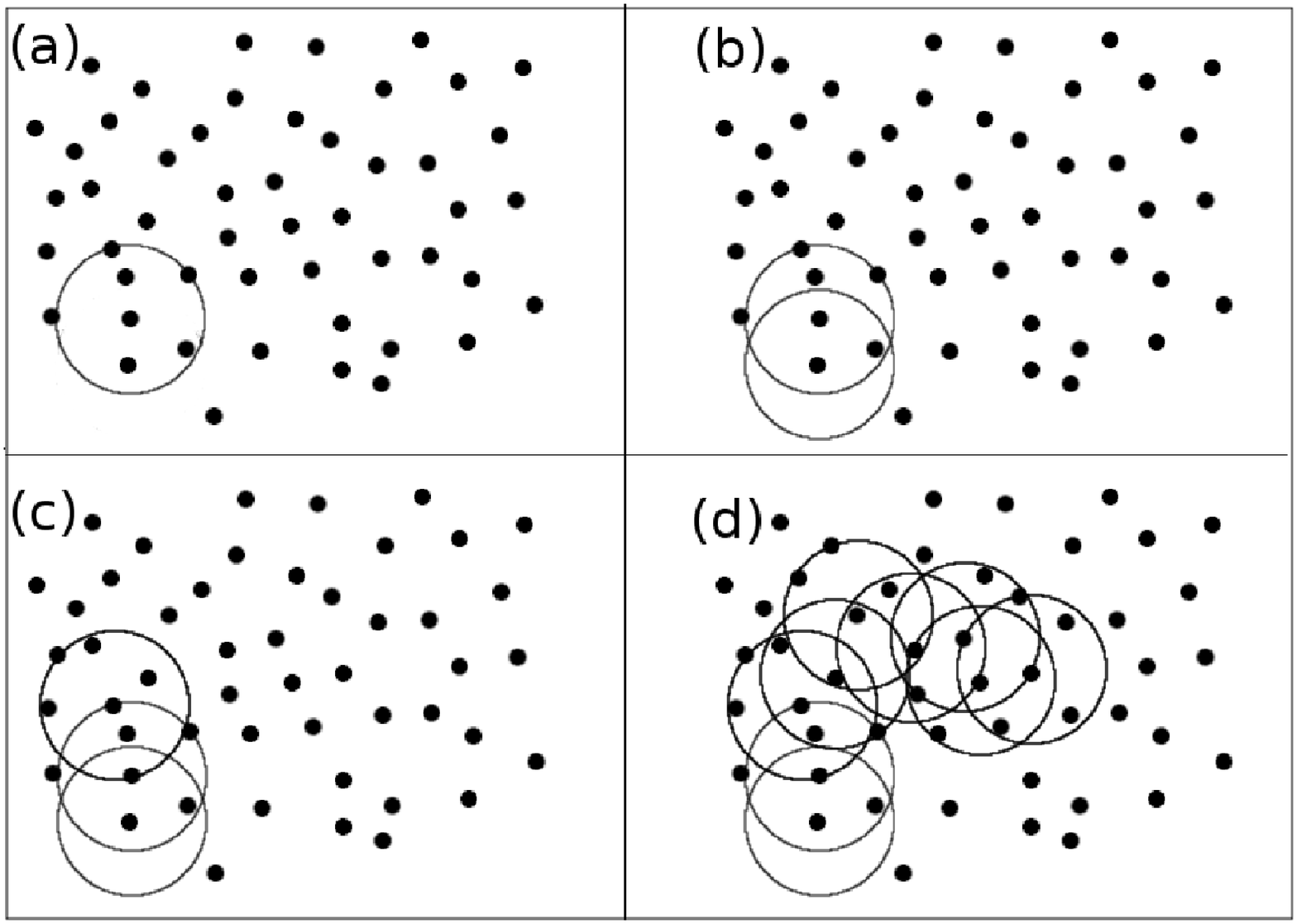}}
\scalebox{1.25}[1.25]{\includegraphics[angle=0,width=60mm]{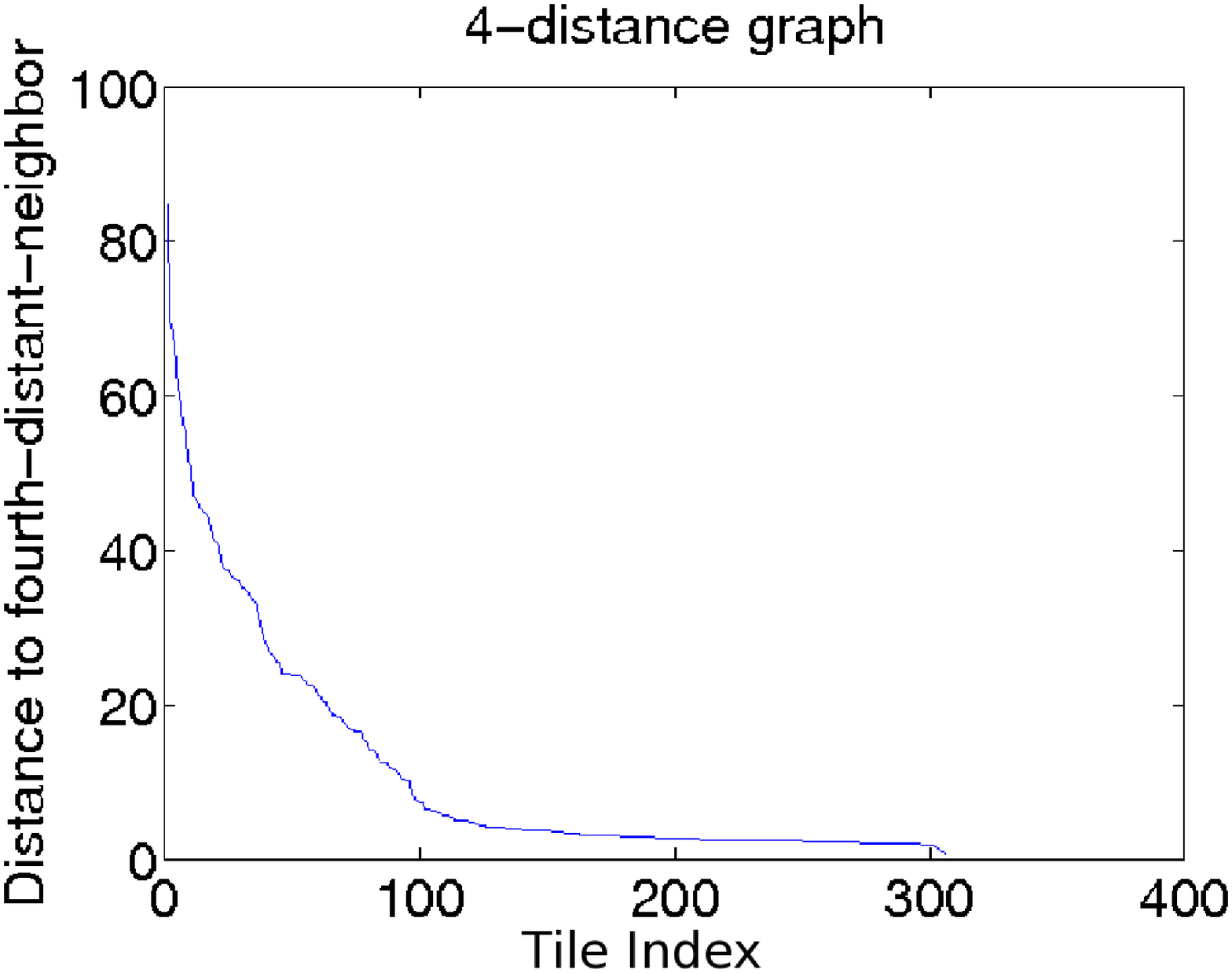}}
\end{center}
\caption{Building clusters from data-points using the density based
clustering algorithm, as discussed in details in
Section~\ref{sec:clustering}. The \textit{left panel} shows the
steps of building a cluster using density based clustering. The
\textit{right panel} shows the 4-distance graph which helps us
determine the neighborhood radius. (Figures derived from concepts
presented in~\cite{ref:dbcluster}).}
\label{fig:density_concept_distance}
\end{figure}

Density based clustering~\cite{ref:dbcluster} facilitates searches for
signals of unknown shape. It does not clutter the output with a list
of numerous noise related clusters that contain just a few significant
data-points. The algorithm looks for neighbors of those points that
have at least a given number of neighboring points within a given
distance on the time-frequency plane, and forms clusters of
data-points that can be related through their common neighbors. Our
implementation of density based clustering algorithm takes two
parameters: minimum neighbor number and neighborhood radius, and it
considers each tile produced by QPipeline as a data-point.

Density based clustering first finds a tile's nearest neighbors, then
that neighbors' neighbors, and so on. In the left panel of
Fig.~\ref{fig:density_concept_distance}, (a) shows data points before
clustering. If the density of data points within a given distance
around a point is above a given threshold to form a cluster, that
point becomes a cluster seed (b). Neighboring data points having a
sufficient number of neighbors are then included in the cluster
(c). This process repeats as long as data points with sufficient
number of neighbors are found (d).

Any clustering algorithm requires measurement of the pairwise
distances between all data points; in our case, the pairwise distance
between all tiles produced by QPipeline.  Unfortunately, the tiles
have varied shapes, which makes measurement of distance between any
pair of tiles somewhat difficult.  We have implemented a simple
distance metric that addresses the issue of varied tile shapes.  For a
pair of tiles with center times $t_1$ and $t_2$, center frequencies
$f_1$ and $f_2$, Q of $q_1$ and $q_2$, and normalized energy of $z_1$
and $z_2$, the distance on the time-frequency plane D is measured from
the following relations:

\begin{equation}
\label{eqn:dist_short}
D = \sqrt{{D_t}^{2} {+} \beta{D_f}^{2}}
\end{equation}

\begin{equation}
\label{eqn:distances_scales}
D_t = \frac{|t_2 {-} t_1|}{\overline{\Delta t}}, D_f = \frac{|f_2 {-} f_1|}{\overline{\Delta f}}
\end{equation}

\begin{equation}
\label{eqn:distances_weights}
\overline{\Delta t} =
\frac{{\Delta t_1}{z_1} {+}
{\Delta t_2}{z_2}}{z_1 {+} z_2},
\overline{\Delta f} =
\frac{{\Delta f_1}{z_1} {+}
{\Delta f_2}{z_2}}{z_1 {+} z_2}
\end{equation}
where $D_t$ is the distance along the time scale, $D_f$ is the
distance along the frequency scale, $\overline{\Delta t} $ is the
scale factor on the time scale, $\overline{\Delta f} $ is the scale
factor on the frequency scale, $\Delta t_1$ and $\Delta t_2$ are
durations, and $\Delta f_1$ and $\Delta f_2$ are bandwidths.

The parameter $\beta$ is a tuning parameter that determines the
relative importance of distance in frequency versus distance in time.
It was determined empirically that when $\beta = 30$, the best results
were achieved for the extended signals considered in
Section~\ref{sec:testing} as measured by the receiver operating
characteristic (ROC) curves presented there.

The mismatch metric in Equation~\ref{eqn:mismatch} can also be
used~\cite{ref:qpipeline} for this purpose.  However, in this initial
study we have chosen to use the simpler approach of Equations
~\ref{eqn:dist_short}, ~\ref{eqn:distances_scales}, and
~\ref{eqn:distances_weights}, and leave the possibility of using
Equation~\ref{eqn:mismatch} for future study.

The minimum neighbor number determines which tiles are to be
considered as potential cluster seeds, and which ones are to be
excluded from the clustering process entirely because they do not have
a sufficient number of neighboring tiles.  A minimum neighbor number
that is too low may result in too many clustering seeds, potentially
producing a large number of clusters, whereas a minimum neighbor
number that is too high may result in the exclusion of too many tiles.

The neighborhood radius indicates the distance the algorithm searches
from a tile in order to find neighboring tiles, and therefore
determines the maximum gap over which the algorithm can cluster.  A
neighborhood radius that is too low may result in a large number of
small clusters, whereas a neighborhood radius that is too high may
result in the creation of one large cluster consisting of all
significant tiles.

The exact numerical value of the neighborhood radius depends on the
minimum neighbor number and the distance metric used.  We have
experimented by varying the minimum neighbor number from 1 to 8, and
found that the best results were achieved with a minimum neighbor
number of 4, as measured by the ROC curves in Section
~\ref{sec:testing}.  The value of the neighborhood radius was then
determined using the 4-distance graph in right panel of
Figure~\ref{fig:density_concept_distance}.  The 4-distance graph shows
the number of tiles on the x-axis whose fourth closest neighbor is
less than the distance shown on the y-axis.  Motivated by the
4-distance plot, we experimented by varying the neighborhood radius
between 3 and 12, and found that a value of 8 produced the best
results based on the ROC curves in Section~\ref{sec:testing}.

\begin{figure}[!t]
\begin{center}
{\includegraphics[angle=0,width=140mm]{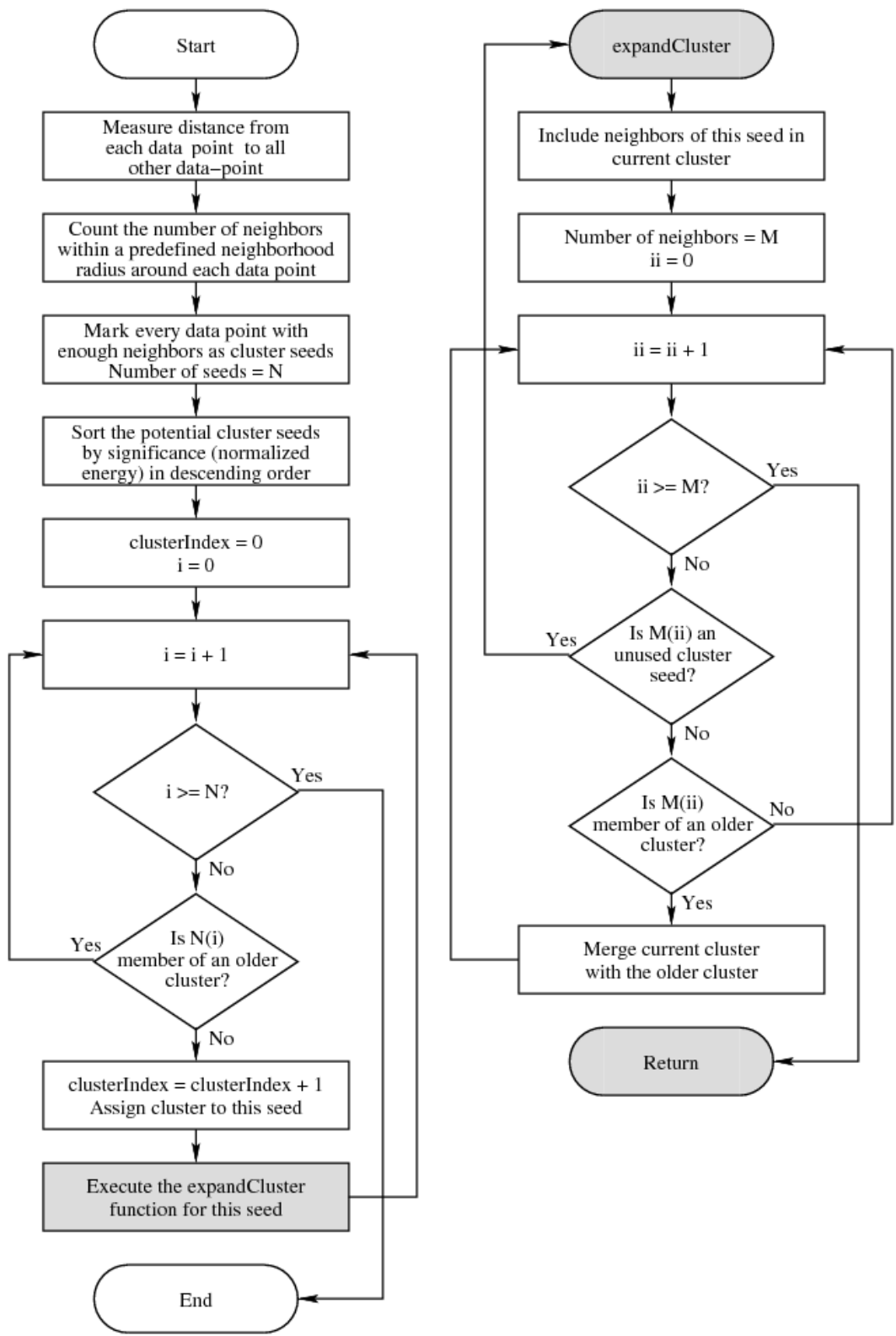}}
\end{center}
\caption{Flowchart of density based clustering algorithm.}
\label{fig:density_clustering_flowchart}
\end{figure}

A flowchart of the density based clustering algorithm is shown in
Figure~\ref{fig:density_clustering_flowchart}.  The main clustering
function first uses the distance function to measure pairwise distance
between all tiles, and then calls the expandCluster function, which
recursively calls itself, to incorporate more tiles into the each
cluster.  The algorithm as presented here was implemented in
MATLAB~\footnote{MATLAB Version $7.2.0.283 (R2006a)$;

$http://www.mathworks.com/access/helpdesk/help/techdoc/rn/f0-68730.html$}.

Clustering starts with the highest energy tile that has a sufficient
number of neighbors, and then proceeds to the next significant tile
that also has a sufficient number of neighboers, and is not already
assigned to a cluster.  If any qualifying member of the current
cluster is found to be already belong to another cluster, the two
clusters are merged.  Thus, regardless of which tile the algorithm
starts clustering from, it will always find the same clusters for a
given set of tiles.  For speed optimization, though, our density based
clustering function starts with the more significant tiles first.

The computational cost of the resulting algorithm is dominated by the
$N^2$ cost of computing the distance between all pairs of tiles, where
$N$ is the number of tiles.  In practice, clustering is applied
separately to the $\sim$1 minute data blocks used by the QPipeline
analysis.  This is more than sufficient to detect clusters up to a few
seconds in duration, the typical limit of gravitational wave burst
searches.  At the typical $\sim$1 Hz single detector false rate, the
resulting computational cost due to clustering is small compared to
that of the rest of the search.

The right panel of Fig.~\ref{fig:example_cont} shows an example
cluster constructed using the proposed density based clustering
algorithm.  It can be seen that the algorithm has clustered together
the most significant part of the previously discussed injection
successfully.  In addition, almost all of the spurious noise tiles
have also been removed. While the high-frequency end of the signal has
been lost, that part contains very little energy, and does not
significantly contribute to the significance of the detected trigger.

%% file: testing.tex
\section{Evaluating performance improvements}
\label{sec:testing}

We have evaluated our implementation of density based clustering by
measuring its effect on the detection of simulated signals injected
into typical single detector LIGO data, and its effect on the rate of
false detections.

\begin{figure}[!ht]
\begin{center}
\includegraphics[angle=0,width=100mm]{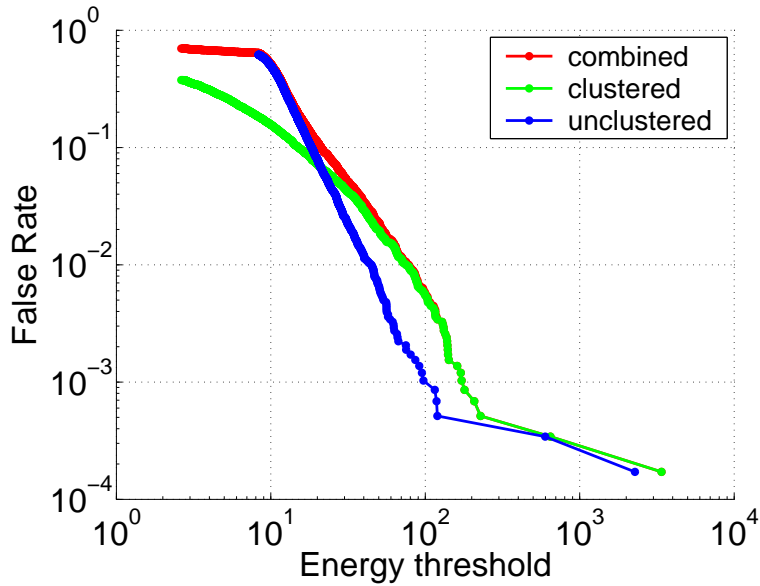}
\end{center}
\caption{The false event rate of the search algorithm as a function of
detection threshold when applied to typical LIGO data.  The trigger
rate is shown for three different trigger sets: unclustered
triggers, clustered triggers, and the union of clustered and
unclustered triggers.}
\label{fig:falserate}
\end{figure}

In order to evaluate the false detection rate, the QPipeline was first
applied to single detector data without injected signals.  This was
performed both with and without clustering.  Since detectable GWB
events are expected to be extremely rare in the few hours of data
considered here, and since we have set our thresholds to yield event
rates up to $\sim$1 Hz and not demanded coincidence between multiple
detectors to reject false events, we can safely identify false events
as those events in a given data stretch whose total normalized energy
exceeded a specified detection threshold.  Three sets of false events
were identified: unclustered events, clustered events, and combined
events formed by the union of unclustered and clustered events.  The
resulting false event rates as a function of detection threshold are
shown in Figure~\ref{fig:falserate}.

\begin{figure}[!ht]
\begin{center}
\begin{tabular}{cc}
\multicolumn{2}{c}{Inspiral signals, SNR 25} \\
\includegraphics[angle=0,width=75mm]{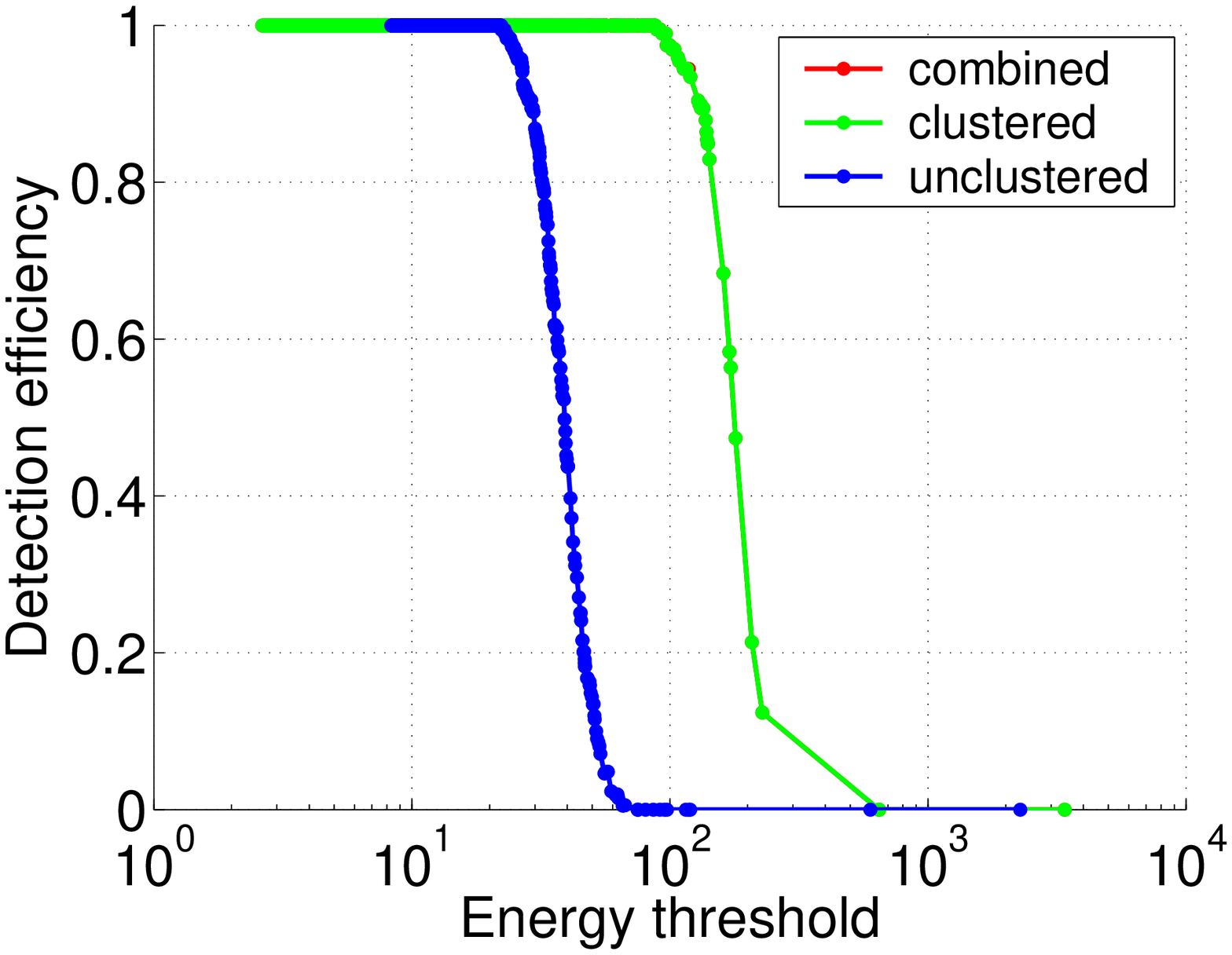} &
\includegraphics[angle=0,width=75mm]{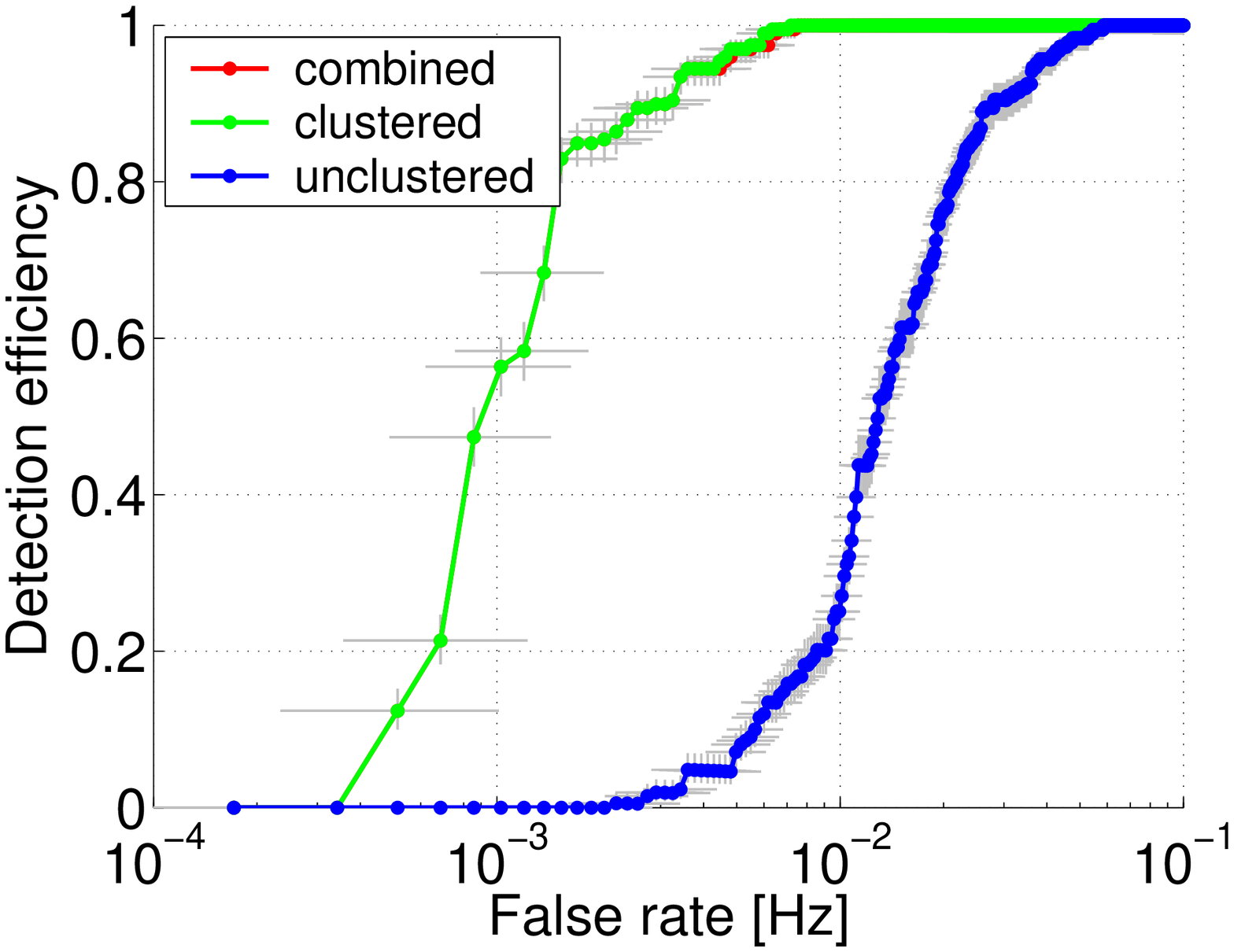} \\
\multicolumn{2}{c}{White noise burst signals, SNR 25} \\
\includegraphics[angle=0,width=75mm]{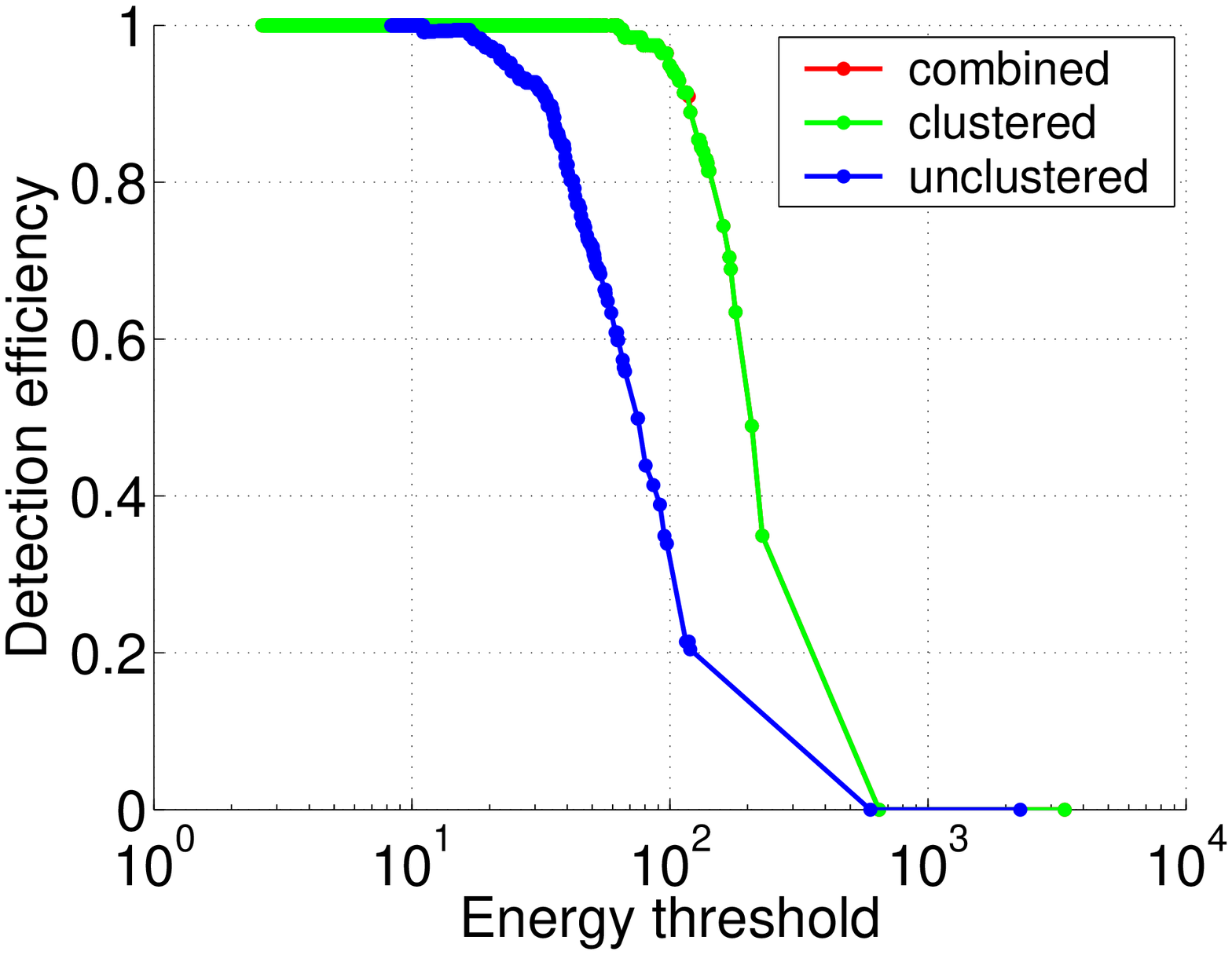} &
\includegraphics[angle=0,width=75mm]{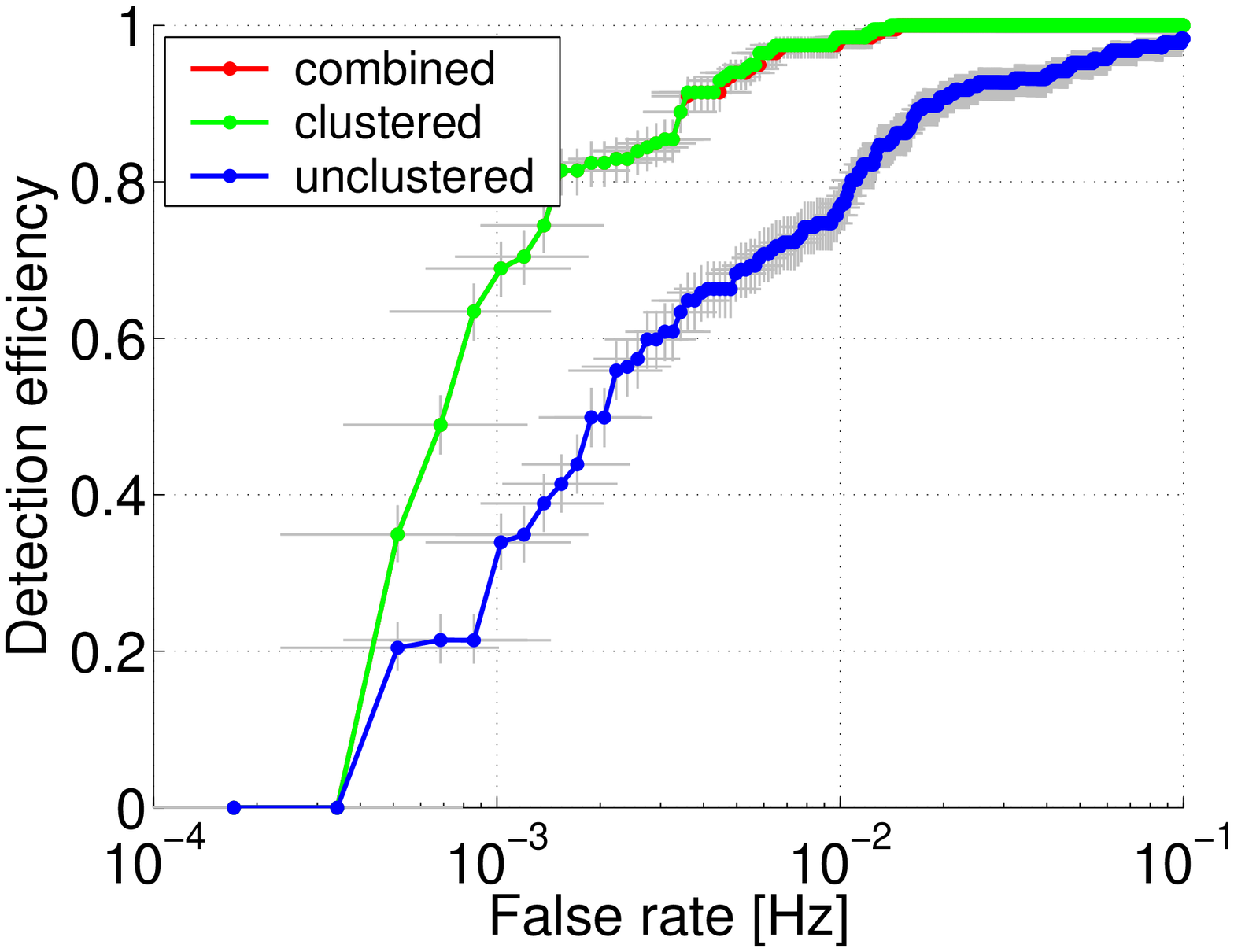} \\
\multicolumn{2}{c}{Sinusoidal Gaussian signals, SNR 10} \\
\includegraphics[angle=0,width=75mm]{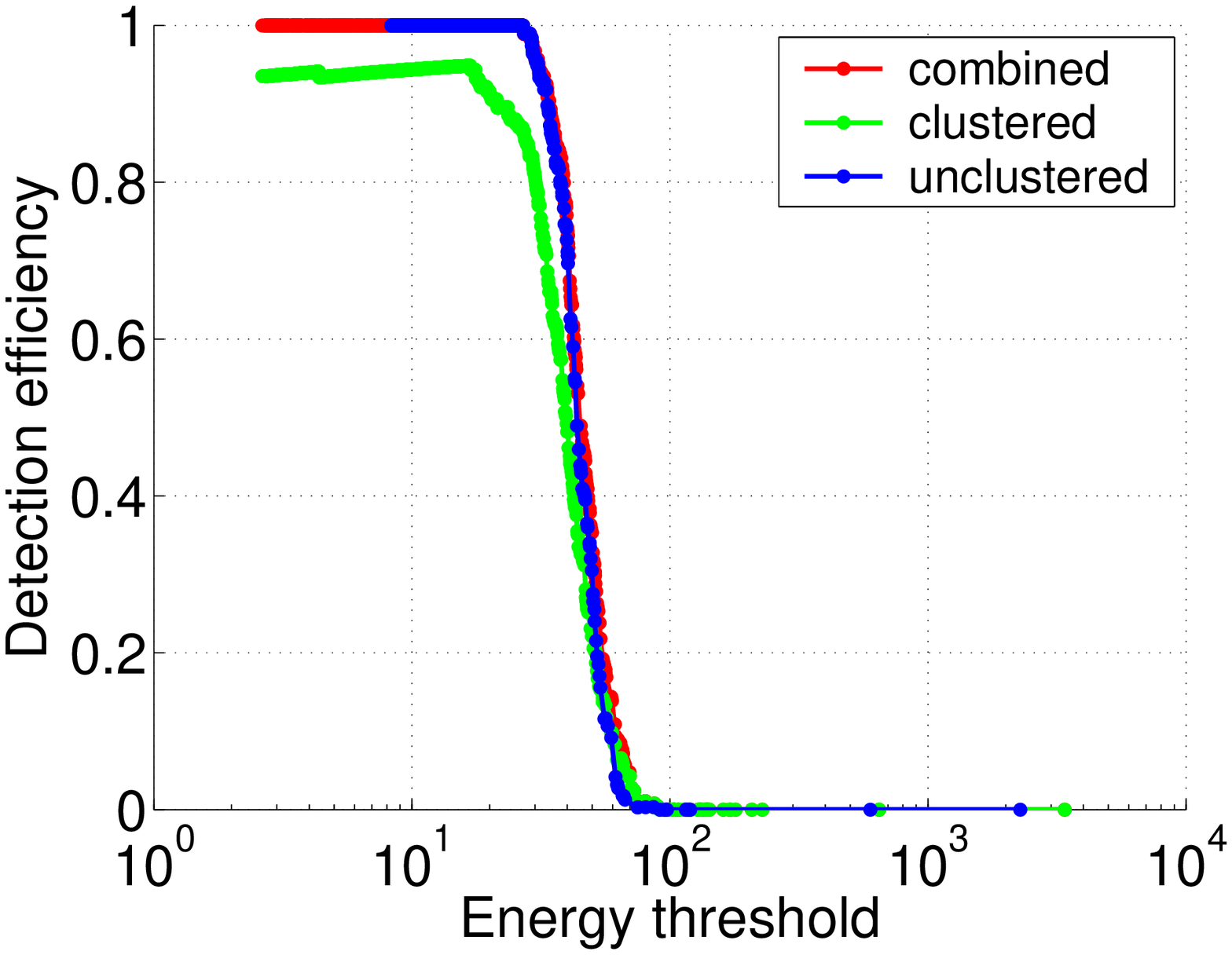} &
\includegraphics[angle=0,width=75mm]{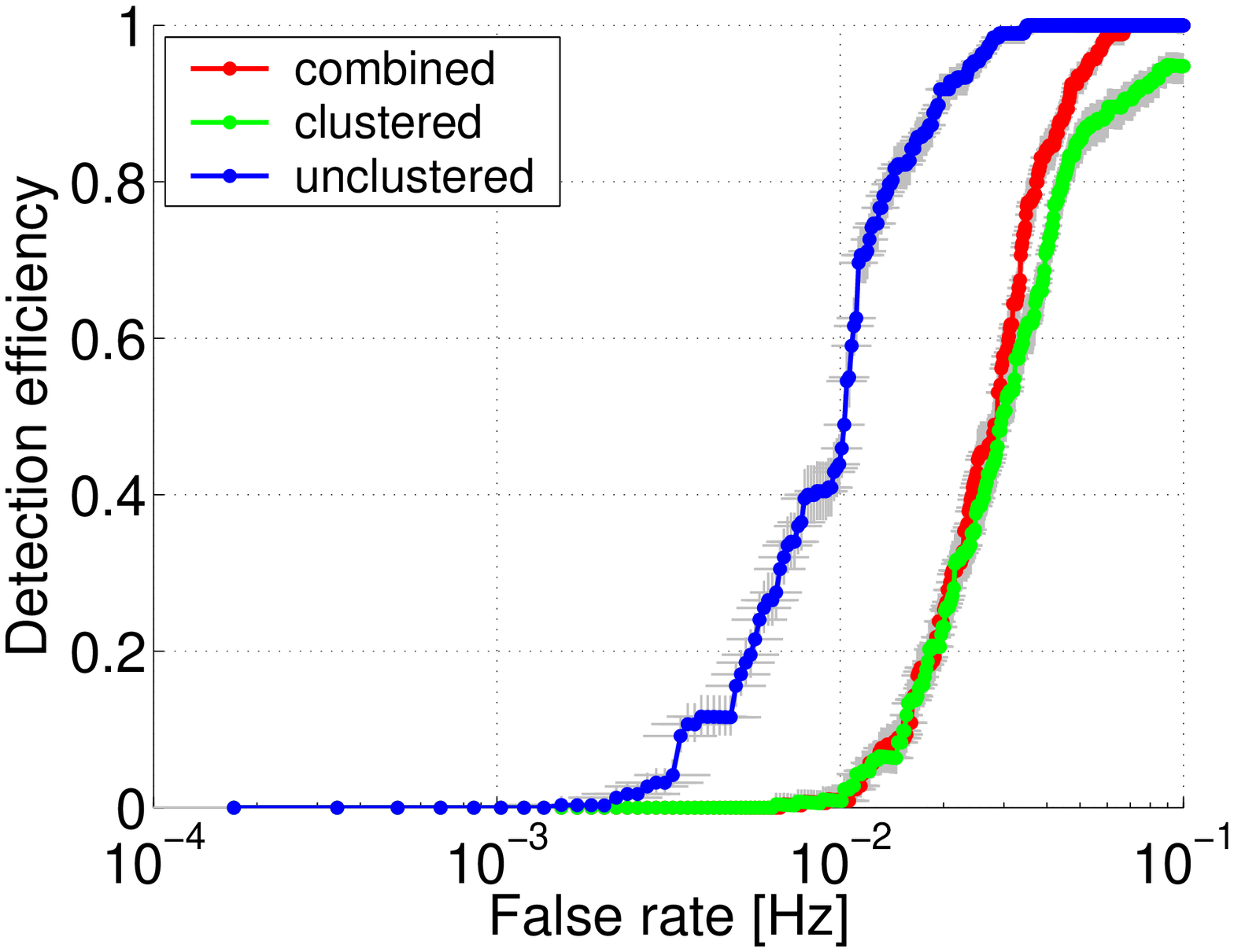} \\
\end{tabular}
\end{center}
\caption{Comparison of the detection efficiency vs. search threshold
(left) and Receiver Operator Characteristics (ROC) (right) of the
search algorithm, with and without clustering, applied to the
detection of simulated inspiral (top), white noise burst (middle),
and sinusoidal Gaussian (bottom) waveforms injected 200 times into
typical LIGO data at fixed SNR.}
\label{fig:roc}
\end{figure}

An adverse effect of using density based clustering is the occasional
rejection of highly localized signals, regardless of the detection
threshold.  This is due to the tendency of density based clustering to
exclude isolated triggers.  This is also evident in
Figure~\ref{fig:roc}, where detection efficiency of sinusoidal
Gaussians does not converge to 100 percent for the case of clustered
triggers.  To overcome this, we have also considered the performance
of a search consisting of the union of both clustered and unclustered
triggers, and compared it with that for clustered triggers and
unclustered triggers only.  The resulting sinusoidal Gaussian combined
detection efficiency for a given energy-threshold is then comparable
to that of the unclustered case, as shown in Figure~\ref{fig:roc}.

Another possible solution to this problem is to reduce the required
number of tiles within the neighborhood radius to zero, permitting
single tile clusters.  Classical hierarchical clustering also provides
an alternative to density based clustering that permits single tile
clusters.  Since the focus of this paper is on improved performance
for signals that are extended in time and/or frequency, we have not
considered either of these alternative choices here.

The lower false event rate observed in Figure~\ref{fig:falserate} for
clustered triggers at low detection thresholds is associated with the
rejection of isolated noise events as described in
Section~\ref{sec:clustering}.  At high detection thresholds, the
opposite is true.  The presence of transient non-stationary
``glitches'' in the data that are extended in time and/or frequency
cause the false event rate of clustered triggers to exceed that of
unclustered triggers.

To evaluate the effect of clustering on the detection of signals, we
next applied the QPipeline to the recovery of simulated signals
injected into the same single detector data.  Again, this was
performed both with and without clustering.  Injections were
identified as detected if a event was observed above the detection
threshold within 1 second of the time of the injected signal.  We
define the detection efficiency as the fraction of injection signals
that were correctly detected, and evaluate this efficiency as a
function of detection threshold for signals injected at a constant
signal to noise ratio.

In order to characterize the performance of density based clustering
for a variety of signal morphologies, we have repeated this analysis
for five different waveform families. They include simple Gaussian
pulses, sinusoidal Gaussian pulses, and the fundamental ring down mode
of perturbed black holes, which represent signals that are highly
localized in the time-frequency plane; and the inspiral phase of
coalescing binary compact objects and band-limited time-windowed white
noise bursts, which are both extended in the time-frequency plane.
Within each waveform family, signals were injected with random
parameters such as time, frequency, duration, bandwidth, mass, etc.

Among non-localized signals, inspirals and white noise bursts
represent two extremes: white noise bursts fill a large time-frequency
region, whereas inspirals, while extended in time and frequency, still
only occupy a small portion of a time-frequency region.  Three of the
waveform families are ad-hoc: simple Gaussian pulses, sinusoidal
Gaussian pulses, and white noise bursts.  Two of the waveform families
were astrophysical: inspirals and ringdowns.  While we are not
designing a search to only target known waveforms such as ringdowns
and inspirals, they are nonetheless also a useful test case because
they are astrophysically motivated and because they can form a basis
for comparison with other existing searches, including matched filter
searches.

On the left side panels of Figure~\ref{fig:roc}, we present the
resulting detection efficiency as a function of detection threshold
for three of the waveform families that we have considered,
representing both ad-hoc and astrophysical, as well as localized and
extended.  On the right side panels of Figure~\ref{fig:roc}, we report
the receiver operator characteristic (ROC) for each waveform, which
combines the measured false rate from Figure~\ref{fig:falserate} with
the detection efficiencies from the left side panels.

The results indicate that for the extended waveforms, such as the
inspiral and noise burst waveforms, clustering increases search
efficiency and significantly improves the resulting ROC by
approximately an order of magnitude in false rate.  The primary reason
for this improved performance is the increase in measured signal
energy due to clustering, which is evident as increased detection
efficiency in the left hand side of Figure~\ref{fig:roc}.

Although clustering provides a marked improvement for the detection of
signals that are extended in time and frequency, Figure~\ref{fig:roc}
indicates that clustering also adversely impacts the performance of
the search for localized waveforms.  In particular, the ROC for
sinusoidal Gaussians is worse by roughly a factor of 3 in false rate
due to the addition of clustering.  The primary cause of this
decreased performance is the higher false event rate, which is due to
the increased significance of detector glitches after clustering, and
is evident in Figure~\ref{fig:falserate}.  For signals that are
extended in time and/or frequency this higher false event rate is more
than compensated by the significant improvement in detection
efficiency, but for more localized signals there is no improvement in
detection efficiency to compensate for the increased false rate.  In
practice, we expect the presence of such detector glitches to be
largely mitigated by the requirement of a coincident and consistent
observation of a gravitational wave in multiple detectors, as well as
the absence of a signal in environmental monitors.  As a result, the
decreased performance for localized signals may also be somewhat
mitigated.

%% file: conclusion.tex
\section{Conclusions}
\label{sec:conclusions}

A density based method for clustering the measurements from
neighboring or overlapping basis functions has been employed to more
efficiently detect GWB signals that are extended in time and/or
frequency, and not well represented by QPipeline's particular choice
of basis.  The method is capable of identifying an arbitrary number of
clusters of arbitrary shape and size, while also rejecting spurious
noise triggers, and does not significantly increase the computational
cost of the overall QPipeline search.

The proposed clustering algorithm itself is not specific to the
QPipeline.  Similar improvements are expected when applied to other
time-frequency searches for gravitational wave bursts. In particular,
the algorithm described here has already been applied to the search
for bursts from the soft gamma repeaters using the flare pipeline
~\cite{ref:flare, ref:flare2, ref:flare3}.  For estimating upper
limits, the flare pipeline initially performed a simple sum over all
frequency bins to measure the total signal energy of a trigger.  The
use of density based clustering instead improved the flare pipeline's
upper limit estimate for 100 ms long white noise bursts in the
frequency band from 64 to 1024 Hz by 42\%.  No improvement was
observed for 22 ms long white noise bursts in the band from 100 to 200
Hz, but such signals are fairly localized in the time-frequency plane.
These results are consistent with our conclusion that density based
clustering is only beneficial when searching for extended signals.

Our implementation of density based clustering is already implemented
as part of the QPipeline, which has now been incorporated into the
$\Omega$-Pipeline~\cite{ref:BurstYr1} for use in future GWB searches.

A number of issues remain open for future investigation.  This paper
has focused only on single detector data.  We have left a study of the
effect of density based clustering on multi-detector GWB searches as a
subject for future investigation.  In Equation~\ref{eqn:dist_short},
we have proposed one possible distance metric.  A study of other
distance metrics, in particular ones based on the mismatch metric of
Equation~\ref{eqn:mismatch} is also possible.  A more in depth study
of hierarchical clustering methods, and comparison with the proposed
density based method, as well as previously proposed
methods~\cite{ref:tfclusters,ref:Tama_cluster,ref:Waveburst} is also
recommended.  Finally, the application of clustering to astrophysical
parameter estimation in the event of a detection also warrants further
investigation.

%% file: acknowledgements.tex
\ack

The authors are grateful for the support of the United States National
Science Foundation under cooperative agreement PHY-04-57528,
California Institute of Technology, and Columbia University in the
City of New York. We are grateful to the LIGO Scientific collaboration
for their support. We are indebted to many of our colleagues for
frequent and fruitful discussion. In particular, we'd like to thank
Albert Lazzarini for his valuable suggestions regarding this project,
and Luca Matone, Zsuzsa M\'{a}rka, Sharmila Kamat, Jameson Rollins,
Peter Kalmus, John Dwyer, Patrick Sutton, Eirini Messeritaki, and
Szabolcs M\'{a}rka for their thoughtful comments on the
manuscript. The authors gratefully acknowledge the LIGO Scientific
Collaboration hardware injection team for providing the data used in
figures~\ref{fig:example} and~\ref{fig:example_cont}. We gratefully
acknowledge the contributions of all the software developers and
programmers in the broader scientific community without whose
incremental achievements over many decades we would not be able to
reach this point where implementing this project has become possible.

The authors gratefully acknowledge the support of the United States
National Science Foundation for the construction and operation of the
LIGO Laboratory and the Particle Physics and Astronomy Research
Council of the United Kingdom, the Max-Planck-Society and the State of
Niedersachsen / Germany for support of the construction and operation
of the GEO600 detector. The authors also gratefully acknowledge the
support of the research by these agencies and by the Australian
Research Council, the Natural Sciences and Engineering Research
Council of Canada,the Council of Scientific and Industrial Research of
India, the Department of Science and Technology of India, the Spanish
Ministerio de Educaciony Ciencia, The National Aeronautics and Space
Administration, the John Simon Guggenheim Foundation, the Alexander
von Humboldt Foundation,the Leverhulme Trust, the David and Lucile
Packard Foundation, the Research Corporation, and the Alfred P. Sloan
Foundation. The LIGO Observatories were constructed by the California
Institute of Technology and Massachusetts Institute of Technology with
funding from the National Science Foundation under cooperative
agreement PHY-9210038. The LIGO Laboratory operates under cooperative
agreement PHY-0107417.  This document has been assigned LIGO document
number LIGO-P070041-01-Z.